\documentclass[12pt]{article}
\usepackage{graphicx}
\usepackage{amsfonts}
\usepackage{amssymb,amsmath}
\usepackage{latexsym}
\usepackage{color}
\usepackage{cite}
\input{colordvi.tex}

\renewcommand{\thefootnote}{\#\arabic{footnote}}

\begin{document}

\renewcommand{\thepage}{\arabic{page}}
\setcounter{page}{1}
\renewcommand{\thefootnote}{\#\arabic{footnote}}

\begin{titlepage}

\begin{center}

\vskip .5in

{\Large \bf Mixed Inflaton and Spectator Field Models \\ after Planck
}

\vskip .45in

{\large
Kari~Enqvist$\,^{1}$ and Tomo~Takahashi$\,^2$
}

\vskip .45in

{\em
$^1$
Department of Physics and
Helsinki Institute of Physics, FIN-00014 University of Helsinki 
\\
$^2$
Department of Physics, Saga University, Saga 840-8502, Japan
}

\end{center}

\vskip .4in

\begin{abstract}

We investigate the possibility that the primordial perturbation has two sources: the inflaton and a spectator field, which is not dynamically important during inflation but which after inflation can contribute to the curvature perturbation. The recent Planck results on the power spectrum and non-Gaussianity  allow us to put constraints on such mixed models. In the generic case, where no specific model for the inflaton or the spectator is assumed, one finds that in the mixed scenario it is possible to have a large trispectrum with $\tau_{\rm NL}\gg (f_{\rm NL})^2$. The constraints on inflation models in the plane of the spectral index and tensor-to-scalar ratio are modified by the presence of a spectator and depend also on
the ratio of the spectator-to-inflaton power $R$. If one chooses the spectator to be the curvaton with a quadratic potential, non-Gaussianities can be computed and imply restrictions on possible values of $R$. We also consider a mixed curvaton and chaotic inflation model and show that even quartic chaotic inflation is still feasible in the context of mixed models.
\end{abstract}
\end{titlepage}

\setcounter{footnote}{0}

\section{Introduction}

With the recent Planck data release, the properties of the primordial perturbation has been
determined with unprecedented accuracy. These include the power as well as the bi- and tri-spectra of the density fluctuations.
The amplitude and the spectral index are now precisely measured and given by
$\ln (10^{10} A_s ) = 3.089^{+0.024}_{-0.027}$ (68\% CL) and $n_s = 0.9603 \pm 0.0073$ (68\% CL), respectively \cite{Ade:2013rta}.
Regarding non-Gaussianity, the non-linearity parameter $f_{\rm NL}$ is now tightly constrained as
$ f_{\rm NL}^{\rm local} = 2.9 \pm 5.7, f_{\rm NL}^{\rm equil} = -42 \pm 75, f_{\rm NL}^{\rm orthog} = -25 \pm 39 $ (68\% CL) for the local, equilateral and orthogonal
types, respectively \cite{Ade:2013tta}. In addition, Planck data also yields a constraint on one of the parameters of the trispectrum
as $\tau_{\rm NL} \le 2800$ (95\% C.L.)  \cite{Ade:2013tta}\footnote{
Although Planck team has not reported a constraint on the other trispectrum parameter $g_{\rm  NL}$, 
there are several works  investigating the constraint on $g_{\rm NL}$ using pre-Planck data 
 \cite{Desjacques:2009jb,Smidt:2010ra,Regan:2010cn,Sekiguchi:2013hza,Giannantonio:2013uqa}.
}.

Theoretically, quantum fluctuations of the inflaton have been the main contender for the explanation of the origin of the density fluctuations.
With the Planck data, many inflation models can now be critically tested. For example, the simplest chaotic inflation model with a quadratic
potential seems to be on the verge of the exclusion because of constraints on the spectral index and the scalar-to-tensor ratio. Chaotic inflation with a purely quartic potential appears to be completely ruled out.  While such simple inflation models may be in trouble, other versions of single field inflation are of course still not ruled out. But there exist yet another simple possibility: the origin of the perturbation could be due to a spectator field. A spectator is a light scalar field with mass much less than the Hubble rate $H_*$ during inflation which
plays no dynamical role during inflation. 
Spectators are irrelevant for inflation, but they may become dynamically important after the end of inflation and
generate some, or even all, of the primordial perturbation. The much studied curvaton \cite{Enqvist:2001zp,Lyth:2001nq,Moroi:2001ct} and the field responsible for
modulated reheating \cite{Dvali:2003em,Kofman:2003nx} are well known examples.

As light fields, spectators are subject to inflationary fluctuations. In the case of modulated (p)reheating, the perturbations of the spectator give rise to a space-dpendent decay rate for the inflaton, thus generating curvature perturbations. The field perturbation of the spectator can also evolve after inflation, as is the case in the curvaton model, thus (partially) decoupling the observed density perturbation from the inflationary era.

In general, the perturbations of the inflaton and the perturbations of the spectator could both contribute to the observed density perturbation; we call such a case a {\em mixed model}\footnote{
This possibility has been  investigated in the context of  both curvatons \cite{Langlois:2004nn,Moroi:2005kz,Moroi:2005np,Ichikawa:2008iq,Suyama:2010uj} 
and modulated reheating \cite{Ichikawa:2008ne}.
}.
For a unified picture of the origin of the primordial perturbation, one should therefore allow for two classes of possible sources: inflaton(s) and spectator(s). One may approach the problem on a general level, but for detailed understanding
one should specify the model of inflation even if spectators are solely responsible for the observed perturbation.
In particular, the details of the inflation model are relevant to the scale-dependence of
the power spectrum and the tensor-to-scalar ratio, on which the Planck data has put severe constraints. These are the issues
we address in the present paper.

The structure of the paper is as follows. In the next section, we summarize the formalism
and give general formulas for the power and bi- spectra to investigate mixed inflaton  and
spectator field models.
In Section~\ref{sec:after_planck}, first we investigate  constrains on a mixed inflaton and spectator field model for a general case, i.e., without 
specifying the spectator model, in the light of recent Planck results. Then, as an example,  we consider a curvaton 
as the spectator field and discuss the status of the models. We consider both generic inflation and more well defined chaotic inflation.
In the final section we summarize and discuss our results.

\section{The formalism}  \label{sec:formalism}

Let us begin  by describing briefly how one can obtain the
curvature perturbation and its scale dependence, or the spectral index $n_s$, and
the tensor-to-scalar ratio $r$ in a mixed model.
We consider a scenario where there is one light
spectator field, such as the curvaton, which together with the inflaton is the
source for the primordial density perturbation.
As limiting cases, the formalism given below also includes
models where
either the spectator or the inflaton is solely responsible for density
fluctuations. More detailed description of the mixed models can be found in \cite{Ichikawa:2008iq,Suyama:2010uj}.

We adopt the $\delta N$ formalism \cite{Starobinsky:1986fxa,Salopek:1990jq,Sasaki:1995aw,Sasaki:1998ug,Lyth:2004gb} 
and denote the inflaton field by $\phi$ and the spectator field by $\sigma$.
The curvature perturbation $\zeta$ reads then
\begin{eqnarray}
\zeta
&=& N_\phi \delta \phi_\ast
+ \frac{1}{2} N_{\phi\phi} (\delta \phi_\ast)^2
+ \frac{1}{6} N_{\phi\phi\phi} (\delta \phi_\ast)^3
+N_\sigma \delta \sigma_\ast
+ \frac{1}{2} N_{\sigma\sigma} (\delta \sigma_\ast)^2
+ \frac{1}{6} N_{\sigma\sigma\sigma} (\delta \sigma_\ast)^3,
\end{eqnarray}
where  we have defined $N_\phi = dN / d \phi$ and so on.
By specifying  explicit models for inflation and for the spectator field,
 we can calculate the observational quantities,
such as the spectral index of the power spectrum $n_s$, the tensor-to-scalar ratio $r$, as well as the
non-linearity parameters $f_{\rm NL}, \tau_{\rm NL}, g_{\rm NL}$, by evaluating the derivatives $N_\phi, N_\sigma$ etc.
The appropriate expressions  are given
below.

We assume that $\phi$ and $\sigma$ are uncorrelated. The power spectrum is then given by
\begin{equation}
\label{definingR }
\mathcal{P}_\zeta (k) = \mathcal{P}_\zeta^{(\phi)} + \mathcal{P}_\zeta^{(\sigma)}
= (N_\phi^2 + N_\sigma^2) \left( \frac{H_\ast}{2\pi} \right)^2
=  (1+ R) \mathcal{P}_\zeta^{(\phi)},
\end{equation}
where $\mathcal{P}_\zeta^{(\phi)}$ and $\mathcal{P}_\zeta^{(\sigma)}$ are
the power spectra for the curvature perturbation generated by $\phi$ and $\sigma$, respectively.
$R$ is the ratio of the power spectrum of the spectator field to that of the inflaton:
\begin{equation}
\label{defRandXi }
R \equiv \frac{\mathcal{P}_\zeta^{(\sigma)}}{\mathcal{P}_\zeta^{(\phi)}},
\end{equation}
In the following, we assume that the total scalar potential is given by the sum of the
contribution from $\phi$ and $\sigma$ as $V(\phi) + U(\sigma)$, consistent with the assumption of
no correlation between the inflaton and the spectator.

By using $R$, 
we may write down simple expressions for the various observables.
The spectral index and the tensor-to-scalar ratio are given by
\begin{equation}
n_s - 1
=
-2 \epsilon + 2 \eta_{\sigma}- \frac{4 \epsilon - 2\eta_\phi}{1+R},
\end{equation}
where $\epsilon$ and $\eta_\phi$ are the slow-roll parameters associated with the inflaton potential. They
are defined in the usual way as
\begin{equation}
\epsilon \equiv \frac12 M_{\rm pl}^2 \left( \frac{V_\phi}{V} \right)^2,
\qquad\qquad
\eta_\phi \equiv M_{\rm pl}^2  \frac{V_{\phi\phi}}{V}.
\end{equation}
 For later convenience, let us also define an additional slow-roll parameter characterizing the third derivative of the inflaton potential by
\begin{equation}
\xi_\phi^2 \equiv  M_{\rm P}^4 \frac{V_\phi V_{\phi\phi\phi}}{V^2}.
\end{equation}
Analogously, we denote  the slow-roll parameter for the spectator field by $\eta_{\sigma}$, which
is defined as
\begin{equation}
\eta_{\sigma} = M_{\rm P}^2  \frac{U_{\sigma\sigma}}{3H_\ast^2}.
\end{equation}
All of these quantities are to be evaluated at the time of horizon crossing.

During inflation, gravitational waves are also produced and its power spectrum is
given by
\begin{equation}
P_T(k_{\rm ref}) = \frac{8}{M_{\rm P}^2} \left( \frac{H_\ast}{2\pi} \right)^2,
\end{equation}
then the tensor-to-scalar ratio can be written as
\begin{equation}
\label{eq:tensor-to-scalar}
r = \frac{16 \epsilon}{1+R}.
\end{equation}

The non-linearity parameters are given by
\begin{eqnarray}
\frac{6}{5} f_{\rm NL}
&=&
\left( \frac{1}{1+R} \right)^2
\left[   \frac{N_{\phi\phi}}{N_\phi^2} + R^2 \frac{N_{\sigma\sigma}}{N_\sigma^2} \right],
\\
\tau_{\rm NL}
&=&
 \left( \frac{1}{1+R} \right)^3
\left[ \left( \frac{N_{\phi\phi}}{N_\phi^2} \right)^2+ R^3 \left( \frac{N_{\sigma\sigma}}{N_\sigma^2} \right)^2 \right],
\\
g_{\rm NL}
&=&
\left( \frac{1}{1+R} \right)^3
\left[ \frac{N_{\phi\phi\phi}}{N_\phi^3}
+ R^3  \frac{N_{\sigma\sigma\sigma}}{N_\sigma^3}  \right].
\end{eqnarray}
Although we do not consider the constraint from $g_{\rm NL}$ in the next section, we
give some expressions for $g_{\rm NL}$  for further reference.

With the slow-roll approximation, the derivatives of the number of $e$-folds $N$
relevant to the inflaton part
can be given by
\begin{eqnarray}
N_\phi
& = &
\frac{1}{\sqrt{2 \epsilon} M_{\rm P}}, \\
N_{\phi\phi}
& = &
\frac{1}{ M_{\rm P}^2} \left( 1 - \frac{\eta_\phi}{2\epsilon} \right), \\
N_{\phi\phi\phi}
&=&
 \frac{1}{\sqrt{2\epsilon} M_{\rm P}^3} \left(
 - \eta_\phi
 - \frac{\xi_\phi^2}{\epsilon}
 + \frac{2 \eta_\phi^2}{\epsilon}
 \right).
\end{eqnarray}
From these equations, one can see that the combinations $N_{\phi\phi} / N_\phi^2$ and
$N_{\phi\phi\phi} / N_\phi^3$ are slow-roll suppressed, on the other hand, the counterpart for
the spectator field is in general larger than these. Thus, in a usual case,  the non-linearity parameters given above
 can be approximated as
\begin{eqnarray}
\frac{6}{5} f_{\rm NL}
& \simeq&
\left( \frac{R}{1+R} \right)^2 \frac{N_{\sigma\sigma}}{N_\sigma^2}, \\
\tau_{\rm NL}
&\simeq&
 \left( \frac{R}{1+R} \right)^3
 \left( \frac{N_{\sigma\sigma}}{N_\sigma^2} \right)^2, \\
g_{\rm NL}
&\simeq&
\left( \frac{R}{1+R} \right)^3 \frac{N_{\sigma\sigma\sigma}}{N_\sigma^3}.
\end{eqnarray}
In particular, one can find the relation between $f_{\rm NL}$ and $\tau_{\rm NL}$ as
\begin{equation}
\label{eq:tauNL_fNL}
\tau_{\rm NL} \simeq  \frac{1+R}{R} \left( \frac65 f_{\rm NL} \right)^2.
\end{equation}
Thus, in a mixed model, if $R$ is very small, 
in other words, if the power spectrum is dominated by the contribution from the 
inflaton, a spectator field can generated a large non-Gaussianity with
$\tau_{\rm NL}$ large even if $f_{\rm NL}$ is very small \cite{Suyama:2010uj}.

For the spectator part, to give explicit expressions for $\eta_\sigma, N_\sigma, N_{\sigma\sigma}$, and higher derivatives,
we need to specify the model. Here we consider the curvaton model, for which the derivatives of $N$
with respect to $\sigma$ are given as follows \cite{Sasaki:2006kq}:
\begin{eqnarray}
N_\sigma
& = &
\frac{2r_{\rm dec}}{3\sigma_\ast}, \\
N_{\sigma\sigma}
& = &
\frac{2 r_{\rm dec}}{9 \sigma_\ast^2} ( 3 - 4 r_{\rm dec} - 2r_{\rm dec}^2),\\
N_{\sigma\sigma\sigma}
&=&
 \frac{8r_{\rm dec}^2}{54 \sigma_\ast^3}
 \left( - 18 + r_{\rm dec} + 20 r_{\rm dec}^2 + 6 r_{\rm dec}^3 \right).
\end{eqnarray}
where we have assumed a quadratic potential for the curvaton, given by
\begin{equation}
\label{quadraticcurv}
U(\sigma) = \frac{1}{2} m_\sigma^2 \sigma^2,
\end{equation}
with $m_\sigma$ being the mass of the curvaton\footnote{
For discussion on curvaton models with non-quadratic potentials, see \cite{Enqvist:2005pg,Enqvist:2008gk,Huang:2008zj,Enqvist:2009zf,Enqvist:2009eq,Enqvist:2009ww}.
}.
We have adopted the sudden decay approximation, and
$\sigma_\ast$ is the field value of the curvaton during inflation. The quantity
$r_{\rm dec}$ is roughly the ratio of curvaton energy $\rho_\sigma$ to radiation energy $\rho_\gamma$ at the time of the curvaton decay and is defined as
\begin{equation}
r_{\rm dec} \equiv
\left.
\frac{3 \rho_\sigma}{4 \rho_\gamma + 3 \rho_\sigma} \right |_{\rm decay}.
\end{equation}
When the curvaton begins its oscillation and decay during radiation dominated era,
$r_{\rm dec}$ can be approximately evaluated as
\begin{equation}
r_{\rm dec} \sim \left( \frac{\sigma_\ast}{M_{\rm P}} \right)^2 \sqrt{\frac{m_\sigma}{\Gamma_\sigma}},
\end{equation}
where $\Gamma_\sigma$ is the decay rate for the curvaton. With the help of the formalism presented in this section,
we may now study what are the Planck implications for the mixed models.

\section{Constraints of mixed models}  \label{sec:after_planck}

As a general comment, we note that when a spectator field is responsible for the density fluctuations,  $R$ should be much greater than 1.
For the curvaton model, with $R$ defined in (\ref{defRandXi }),  one finds
\begin{equation}
\label{eq:R_curvaton}
R  =  \frac89 \epsilon \left( \frac{M_{\rm P}}{\sigma_\ast} \right)^2 r_{\rm dec}^2.
\end{equation}
From this formula, we can see that the larger value of $\epsilon$, the larger the curvaton contribution.
In other words, in small field inflation models, the curvaton tends to give small contribution to density perturbations.
We should also note that when $R \gg 1$, the tensor-to-scalar ratio becomes very small.
Since some large field inflation models are excluded due to too large tensor-to-scalar ratio
such as chaotic inflation models with a higher polynomial potential,
for such models, larger values of $R$ can alleviate the inflation model.
However, on the other hand, as pointed out in the Introduction, Planck now precisely measures the spectral index
which implies a stringent constraint on the mixed model.
When $R$ is large, the spectral index is given by
\begin{equation}
n_s -1 \simeq - 2 \epsilon.
\end{equation}
Thus, to have a value for $n_s$ consistent with the observations, one has to invoke a large value for $\epsilon$\footnote{
More precisely, the spectral index has a contribution from the (effective) mass for a spectator field and is given by
\begin{equation}
n_s -1 = -2\epsilon + 2 \eta_{\sigma}.
\end{equation}
For discussion on the spectral index in a spectator field model in the light of Planck result,
see \cite{Kobayashi:2013bna}.
}. Therefore, when one considers a mixed model, or even a pure spectator model,
one needs to take into account the constraints from both the spectral index and the tensor-to-scalar ratio.

\subsection{Generic inflation}
\begin{figure}[htbp]
  \begin{center}
    \resizebox{100mm}{!}{
    \includegraphics{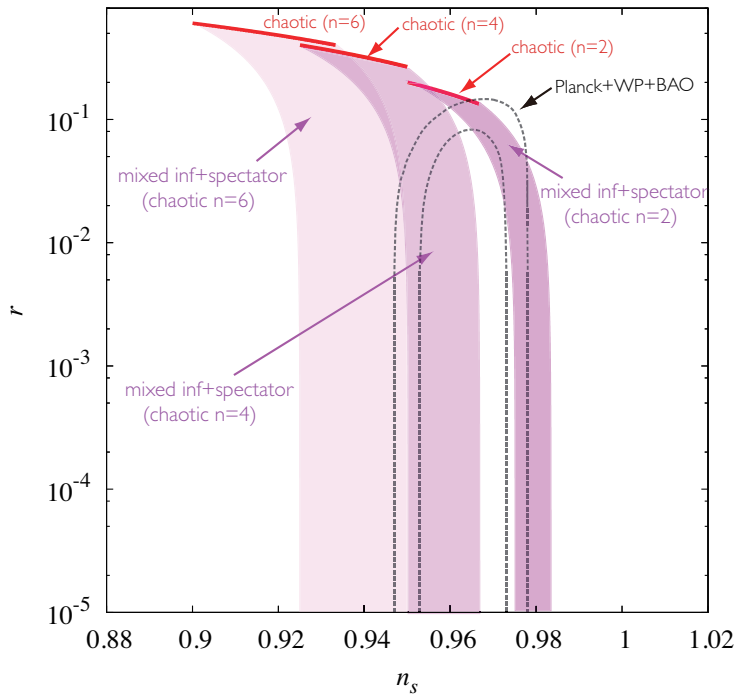}
}
  \end{center}
  \caption{Constraint from Planck+WP+BAO  in the $n_s$--$r$ plane \cite{Ade:2013rta}.
  Predictions for mixed inflaton and spectator field models are also shown.}
  \label{fig:ns_r}
\end{figure}

Let us first discuss the constraints on spectator field models for the general case (i.e., without specifying the inflation model)
with focus on
the spectral index $n_s$ and the tensor-to-scalar ratio $r$.
But, before discussing the constraint, in Fig.~\ref{fig:ns_r}, we show the constraint from Planck in the $n_s$--$r$ plane, 
with predictions of mixed inflaton and spectator field models assuming chaotic inflation,
from which we can have a feeling how such models are constrained from Planck data.
A number of inflation models are now ruled out by virtue of the tension between their predictions for $n_s$ and $r$ and the actual observation.
For mixed models in general, the constraints  in the $n_s$--$r$ plane depend on the  slow-roll parameters  $\epsilon, \eta_\phi$ for
the inflaton sector, and on the ratio of the spectator-to-inflaton power, or  $R$. Depending on the inflation model at hand, some spectators may help them to be resurrected, or alternatively, to be ruled out. For example, when one assumes a quadratic chaotic inflation model,
some cases with large $R$ (corresponding to small $r$ region) turns out to be excluded by
the Planck constraint on $n_s$ (see Fig.~\ref{fig:ns_r}).

\begin{figure}[htbp]
  \begin{center}
    \resizebox{150mm}{!}{
    \includegraphics{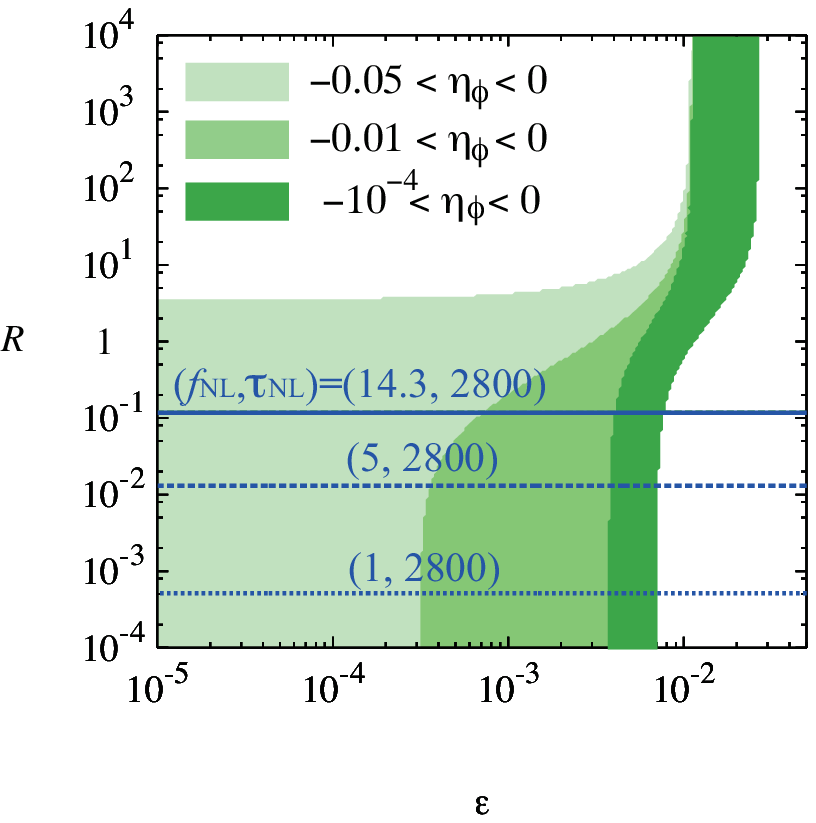}
     \hspace{5mm}
        \includegraphics{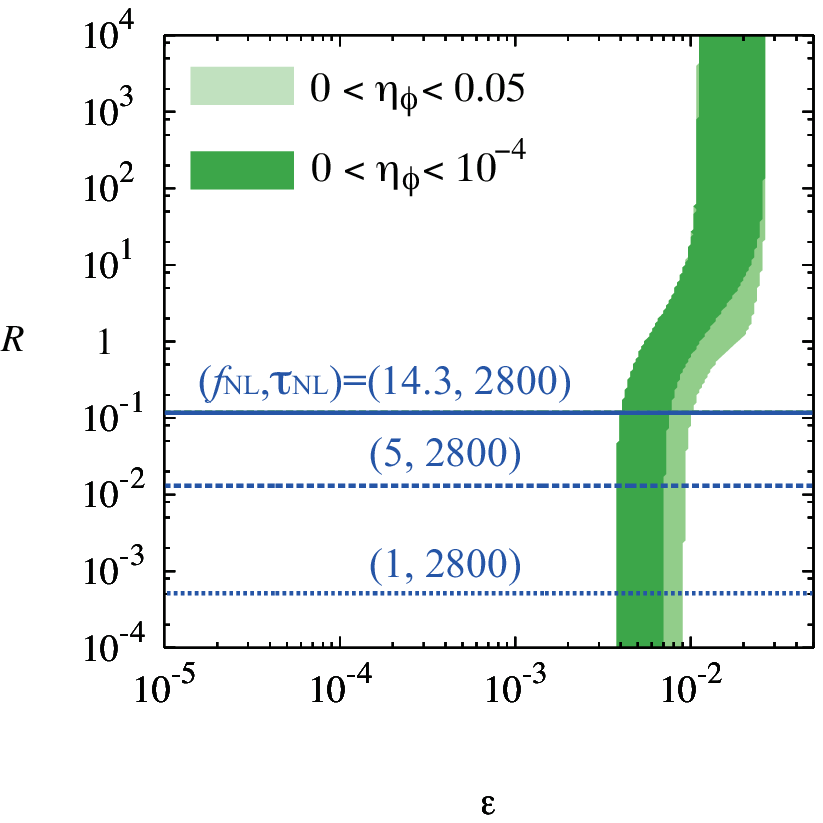}
}
  \end{center}
  \caption{Region allowed in mixed inflaton and spectator models by the constraints on $n_s$ and $r$ from Planck+WP+BAO
  on the $\epsilon$--$R$ plane (shaded region). No specific inflation model is assumed, and
  $\eta_\phi$ is allowed to vary within the ranges indicated in the figures. 
  In the left and right panels, negative and  positive values for $\eta_\phi$ are assumed, respectively.}
   For reference, we show contours of $(f_{\rm NL}, \tau_{\rm NL})=(14.3, 2800), (5, 2800)$ and $(1, 2800)$, below which
   the value of $R$ is excluded by the constraint from $\tau_{\rm NL}$ for the fixed $f_{\rm NL}$.
  \label{fig:eps_R}
\end{figure}

However, the Planck data on $n_s$ and $r$ allows us to put some constraints on mixed models without specifying the inflaton or spectator model.
In Fig.~\ref{fig:eps_R}, we show the  2$\sigma$ allowed region (shaded) in the $\epsilon$--$R$ plane, which is a translation of the constraints on $n_s$ and $r$.
To obtain $n_s$, we here assume that the mass of the spectator should be much smaller than the Hubble parameter during
inflation, and set $\eta_\sigma =0$. In order to make the constraint general, we also freely vary the value of $\eta_\phi$
within the ranges indicated in the Fig.~\ref{fig:eps_R},
assuming negative and positive values for $\eta_\phi$, in the left and right panels, respectively.
Here $R$ is treated as free parameter; thus Fig.~\ref{fig:eps_R} is applicable to any general spectator field model.
When $R \ll 1$, i.e., the inflaton gives a dominant contribution to the power spectrum,
by tuning the value of $\eta_\phi$, a model can be made consistent with Planck data except for the region with
large $\epsilon$, which can occur for a negative $\eta_\phi$. Hence the allowed region is larger for the case with a negative $\eta_\phi$.
On the other hand, an inflation model with positive $\eta_\phi$ is difficult to realize in a successful spectator
field model due to the constraint on $n_s$.

Non-linearity parameters such as $f_{\rm NL}$ and $\tau_{\rm NL}$ cannot be determined
by $\epsilon$ and $R$ alone, but we need to know second order quantities that are usually highly model-dependent. However,
 given the value of $f_{\rm NL}$, we can draw contours of $\tau_{\rm NL}$ by assuming that the value of $f_{\rm NL}$
is fixed by tuning some of the parameters in a model. For example, in the curvaton case, this corresponds to a fixed $r_{\rm dec}$
while keeping $R$ unchanged, which can be achieved by a tuning of $\sigma_\ast$, as can be seen in Eq.~\eqref{eq:R_curvaton}.
In Fig.~\ref{fig:eps_R}, we show  contours of $\tau_{\rm NL} = 2800$ for some fixed values of $f_{\rm NL}$, below
which the parameter region is excluded by Planck $\tau_{\rm NL}$ constraint.
As discussed in the previous section, as $R$ is decreased, $\tau_{\rm NL}$ is enhanced, which in some cases gives a more stringent
constraint.

\begin{figure}[htbp]
  \begin{center}
    \resizebox{150mm}{!}{
    \includegraphics{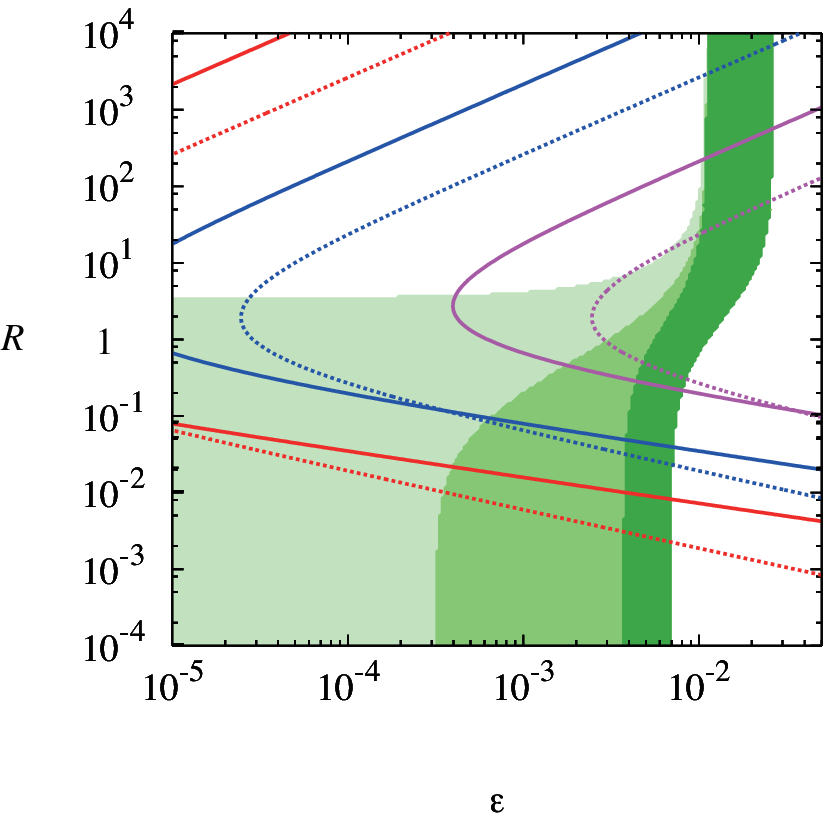}
    \hspace{5mm}
        \includegraphics{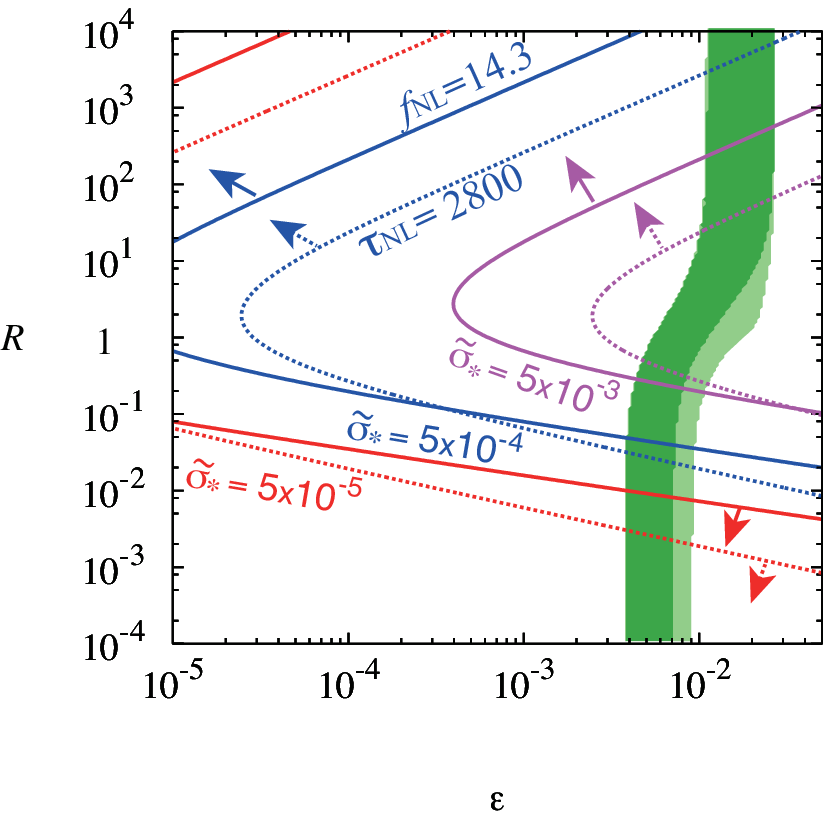}
}
  \end{center}
  \caption{
 Contours of $f_{\rm NL}=14.3$ (solid line) and $\tau_{\rm NL} = 2800$ (dotted line) 
  for the curvaton model with $\tilde{\sigma}_\ast (\equiv \sigma_\ast / M_P) =  5 \times 10^{-3}$ (purple),  $ 5 \times 10^{-4}$ (blue)
  and $5 \times 10^{-5}$ (red), superimposed on a general constraint on $R$ and $\epsilon$ from $n_s$ and $r$ (the same as Fig.~\ref{fig:eps_R}.). Arrows in the figure indicate the direction to which the values of $f_{\rm NL}$ and $\tau_{\rm NL}$ decreases, i.e., 
  correspond to the allowed region given Planck non-Gaussianity constraints.
   }
  \label{fig:eps_R_1}
\end{figure}

Up to now,  our discussion has been model-independent.
However, often many of the model parameters are related to each other in a complicated way. This can lead to additional constraints.
To be specific, let us focus on the curvaton model with the potential given by (\ref{quadraticcurv}). We still treat inflaton sector in a model-independent way.
In Fig.~\ref{fig:eps_R_1}, we also show
the contours of $f_{\rm NL}=14.3$ and $\tau_{\rm NL}=2800$, which both correspond to 2$\sigma$ upper limit from Planck,
  for the curvaton model superimposed for several fixed values of the initial curvaton field value $\sigma_\ast$.
As mentioned above, to be able to find values of $f_{\rm NL}$ and $\tau_{\rm NL}$,
we need to specify the spectator field model explicitly. This is because in order to obtain non-Gaussianities, one needs to compute $N_{\sigma\sigma}$, which
cannot be deduced from $R$ alone.

For the curvaton model, $f_{\rm NL}$ is given by \cite{Lyth:2005du,Lyth:2005fi,Sasaki:2006kq}
\begin{equation}
\frac56 f_{\rm NL}
= \frac{3}{2 r_{\rm dec}} \left( 1 + \frac{\sigma_{\rm osc} \sigma_{\rm osc}^{''}}{\sigma_{\rm osc}^{'2}}  \right) - 2 - r_{\rm dec},
\end{equation}
where $\sigma_{\rm osc}$  represents the curvaton field value at the start of its oscillation; a prime
denotes a derivative with respect to $\sigma_\ast$, i.e., $\sigma_{\rm osc}^{'} = d \sigma_{\rm osc} / d \sigma_\ast$.
We have included a term with $\sigma_{\rm osc}$ for a generality. However, $\sigma_{\rm osc}^{''}$ vanishes
for the curvaton with a quadratic potential assumed here.

Recall that the Planck data yields constraints for these non-linearity parameters as $ -8.9 < f_{\rm NL} < 14.3$ and $\tau_{\rm NL} < 2800$
(95\% C.L.). Thus some of the parameter space is excluded by the non-Gaussianity constraints.
It is interesting to note that, in some cases, a third order non-Gaussianity parameter $\tau_{\rm NL}$ can
give a more stringent constraint than $f_{\rm NL}$. This is because $f_{\rm NL}$ can be suppressed
by a small value of $R$, in which case $\tau_{\rm NL}$ will be
enhanced relative to $(f_{\rm NL})^2$, as is evident in Eq.~\eqref{eq:tauNL_fNL}.

\subsection{Mixed curvaton and chaotic inflation}

Let us now study a specific spectator model, the curvaton model, assuming a well defined inflation model, which we take to be chaotic
inflation. It has the potential
\begin{equation}
V(\phi) = \lambda M_{\rm P}^4  \left( \frac{\phi}{M_{\rm P}} \right)^n.
\end{equation}
In this model, the slow-roll parameters, the spectral index and the tensor-to-scalar ratio are
given as
\begin{eqnarray}
&&
\epsilon
= \frac{n}{4 N_{\rm inf}},
\qquad
\eta_\phi
=  \frac{n-1}{2 N_{\rm inf}}
= \frac{2(n-1)}{n} \epsilon,\\
&&
n_s -1 = - \frac{n+2}{2N_{\rm inf}},
\qquad
r = \frac{4n}{N_{\rm inf}}.
\end{eqnarray}
Although chaotic inflation models with $n \ge 3$ as the sole source for the curvature perturbation has now been excluded by the Planck data because of the constraints on $n_s$ and $r$, it is interesting to note that within the context of a mixed model,
chaotic inflation can  still be viable, as can be seen from Fig.~\ref{fig:ns_r}.

\begin{figure}[htbp]
  \begin{center}
    \resizebox{150mm}{!}{
    \includegraphics{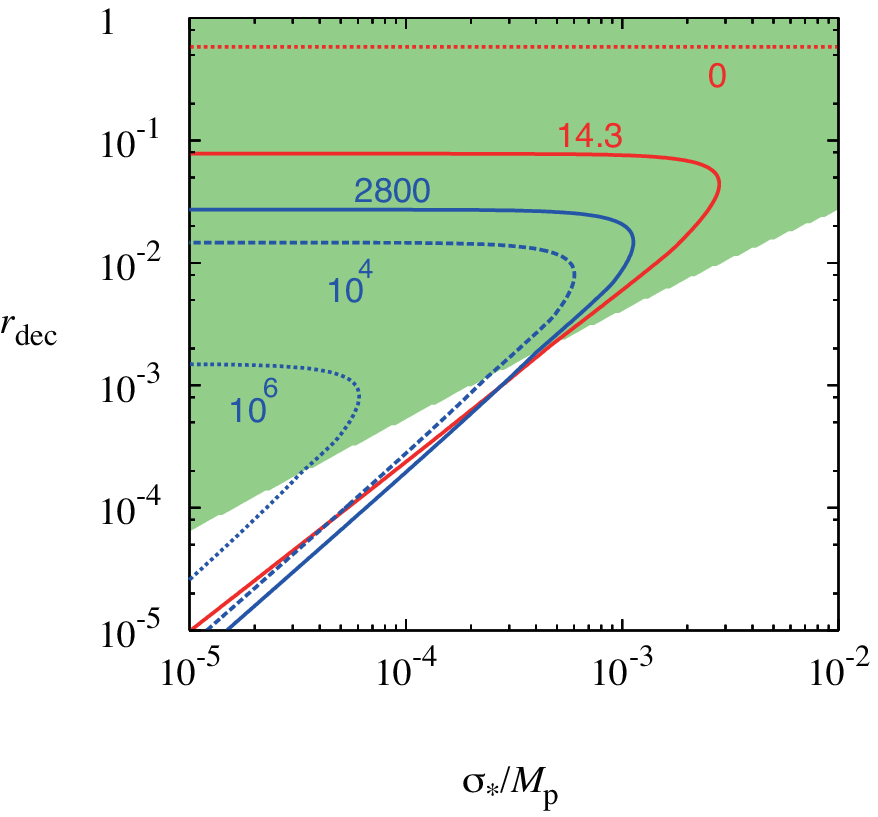}
     \includegraphics{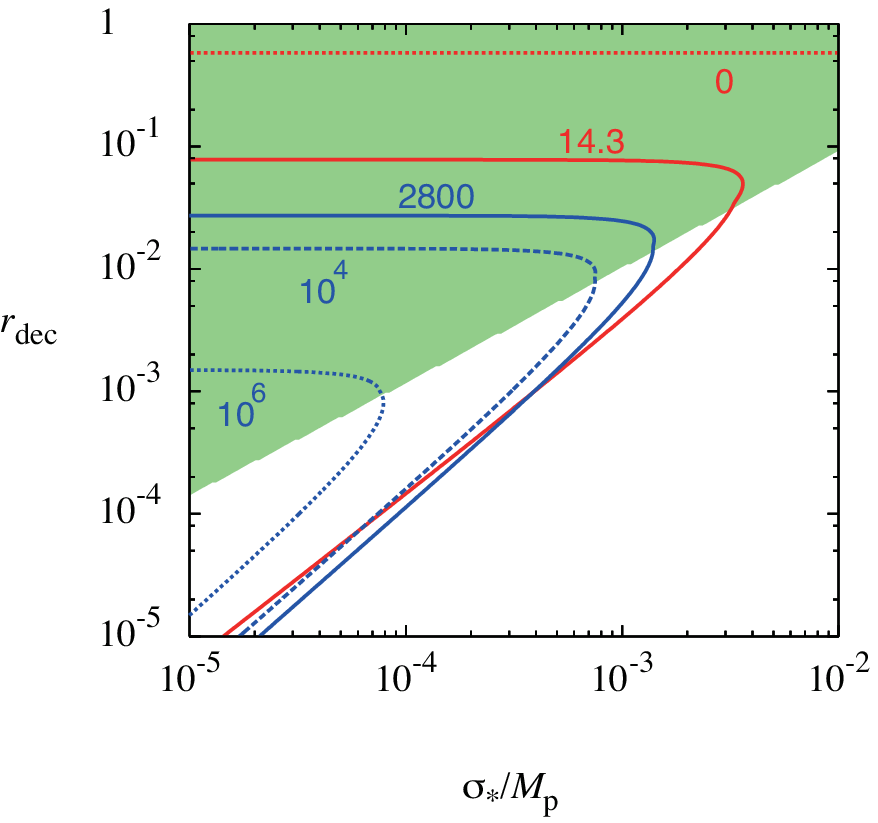}
}
  \end{center}
  \caption{The region allowed by the constraints on $n_s$ and $r$ from Planck
  in the plane of the curvaton parameters $\sigma_\ast$ and $r_{\rm dec}$ (shaded region) for
  a mixed curvaton and chaotic inflation model with a quadratic (left) and quartic (right) potential.
  We let $N_{\rm inf}$  vary as $40 < N_{\rm inf} < 60$.   Contours of $f_{\rm NL}$ (red lines) and $\tau_{\rm NL}$ (blue lines) 
  are also shown, whose values are indicated in the figure.
  }
  \label{fig:sigma_rdec}
\end{figure}

In Fig.~\ref{fig:sigma_rdec},  we show the constraints on $\sigma_\ast$ and $r_{\rm dec}$ as implied by the Planck data.
 The 2$\sigma$ allowed region is shaded green, which follows from the observational constraints on $n_s$ and $r$.
To take into account the uncertainty of the thermal history after inflation, we allow
the number of $e$-folds to vary between $40 < N_{\rm inf} < 60$.  Contours of $f_{\rm NL}$
are also shown in the figure. To compute $f_{\rm NL}$ and $\tau_{\rm NL}$, we fix the value of $\epsilon$ by taking $N_{\rm inf}=60$
for reference. Other than $\epsilon$, there is no other ambiguity for the prediction of $f_{\rm NL}$ and $\tau_{\rm NL}$
once $r_{\rm dec}$ and $\sigma_\ast$ are fixed.

\section{Summary}  \label{sec:summary}

Recent Planck data has put stringent constraints on the nature of the primordial density
perturbations, which enable us to study the details of their origins.
In this paper, we have considered a scenario where the primordial perturbation is a mixture of two sources: the inflaton field and a spectator field, which is a scalar that is dynamically unimportant  during inflation. However, once inflation is over, spectators may contribute to the curvature perturbation. We investigated the mixed spectator-inflaton model using the the Planck results on
the spectral index $n_s$, the tensor-to-sclar ratio $r$,  the non-Gaussianity parameters $f_{\rm NL}$ and $\tau_{\rm NL}$. We first considered the generic case, where the inflaton model and the spectator model is not specified. We noted that although after Planck  $f_{\rm NL}$ is known to be small, in mixed models it is still possible to have large trispectrum with $\tau_{\rm NL}\gg (f_{\rm NL})^2$.

The Planck data rules out some inflation models, such as monomial chaotic inflation $V\propto \phi^n$
with $n>2$, but as we pointed out, the situation changes when one also allows for a spectator field to source the perturbation. In that case the
constraints in the $n_s$--$r$ plane also depend on the ratio of the spectator-to-inflaton power, which we denoted as $R$ and defined in (\ref{definingR }).
The Planck data circumscribes an allowed region in the plane of the slow roll-parameter $\epsilon$ and $R$, which could be further constrained by $\tau_{\rm NL}$. This holds independently of the details of the inflaton or the spectator model. We also noted that the underlying inflation model is relevant even when
one considers pure spectator field models.

More stringent restrictions can be obtained if one assumes a specific spectator model. We chose to consider the curvaton model with a quadratic potential while keeping the inflaton sector arbitrary. In that case, one may compute  $f_{\rm NL}$ and  $\tau_{\rm NL}$ explicitly. 
They were shown to constrain possible values of $R$, as depicted in Fig.~\ref{fig:eps_R_1}. We also discussed a mixed curvaton-chaotic inflation model. We showed that in the context of mixed models even quartic chaotic inflation is still feasible. 
Since more cosmological data relevant for the power spectrum, such  as information about CMB polarization from Planck and other experiments, is expected soon, the allowed parameter space of the mixed models may be expected to shrink further in the near future.

\section*{Acknowledgments}

TT would like to thank the Helsinki Institute of Physics for the hospitality
during the visit, where this work was initiated.
The work of TT is partially supported by the Grant-in-Aid for Scientific
research from the Ministry of Education, Science, Sports, and
Culture, Japan, No.~23740195. KE is supported by the Academy of Finland grants 1263714 and 1218322.

\end{document}